\documentclass{osa-article}

\journal{oe}

\articletype{Research Article}
\usepackage{braket}
\usepackage{tabularx}
\newcolumntype{C}[1]{>{\centering\arraybackslash}p{#1}}
\usepackage{multirow, booktabs}
\usepackage[font=scriptsize]{caption}
\usepackage{subcaption}
\usepackage{cite}

\begin{document}

\title{Strain dependence of photoluminescence and circular dichroism in transition metal dichalcogenides: a \textit{k} {\small\textbullet} \textit{p} analysis}

\author{Shahnaz Aas,\authormark{1} Ceyhun Bulutay,\authormark{1,*}}

\address{\authormark{1}Department of Physics, Bilkent University, 06800, Bilkent, Ankara, Turkey}

\email{\authormark{*}bulutay@fen.bilkent.edu.tr} 


\begin{abstract}
Within a two-band $k\cdot p$ method we analyze different types of strain for the $K$ valley optical characteristics of a 
freestanding monolayer MoS$_2$, MoSe$_2$, WS$_2$ and WSe$_2$. We predict that circular polarization selectivity for energies 
above the direct transition onset deteriorates/improves by tensile/compressive strain. Wide range of available strained-sample 
photoluminescence data  can be reasonably reproduced by this simple bandstructure combined with accounting for excitons at 
a variational level. According to this model strain impacts optoelectronic properties through its hydrostatic component, 
whereas the shear strain only causes a rigid wavevector shift of the valley. Furthermore, under the stress loading of flexible 
substrates the presence of Poisson's effect or the lack of it are examined individually for the reported measurements.
\end{abstract}

\section{Introduction}
Transition metal dichalcogenides (TMDs) possess direct optical gap together with mechanical flexibility up to 10\% range \cite{bertolazzi11} which enables wide strain tunability of their optoelectronic properties \cite{akinwande14, roldan15}. As a consequence, the associated body of literature is rapidly growing, while a number of milestones have been reached. The tuning of the electronic structure by applying a uniaxial tensile bending to monolayer MoS$_2$ on flexible substrates has been demonstrated by several groups within a short time span \cite{he_experimental_2013, conley_bandgap_2013, zhu_strain_2013, castellanos-gomez_local_2013, tonndorf13,  hui_exceptional_2013, sercombe13}. For a suspended monolayer MoS$_2$ membrane, Lloyd {\em et al.} showed the continuous and reversible tuning of the optical bandgap over an ultralarge range of applied biaxial strain \cite{lloyd_band_2016}. More recently, deterministic two-dimensional array of quantum emitters from thin TMDs due to a localized strain pattern is achieved that becomes instrumental to construct scalable quantum architectures \cite{branny16, palacios-berraquero_large-scale_2017}. Additional experimental \cite{plechinger_control_2015, li_optoelectronic_2015, desai_strain-induced_2014, yang_tuning_2015, wang_strain-induced_2015, zhang16, schmidt_reversible_2016, island_precise_2016} as well as theoretical \cite{peelaers12, rostami_theory_2015, maniadaki_strain_2016, feierabend_impact_2017} studies substantiated strain as a viable control mechanism for these two-dimensional materials. 

Our aim in this work is to consolidate accumulating experimental photoluminescence (PL) data on strained TMD samples with the aid of a simple 
$K$-valley-specific two-band $k\cdot p$ theory. Particularly, for the measurements performed by uniaxial bending of flexible substrates, this analysis
can reveal the extend of Poisson's contraction over the TMD layer for each individual case. It also governs the circular dichroism which refers 
to the helicity-selective optical absorption \cite{cao_valley-selective_2012}, and in which way it can be altered by strain. 
\section{Theory}

\subsection{Two-band strained $k\cdot p$ approach for TMDs}\label{sec: k.p approach}
The conduction and valence bands of TMDs around the direct bandgap at the $K_\pm$ valleys can be represented by a two-band 
basis $\left\{\Ket{\phi_c},\Ket{\phi^\pm_v}\right\}$ which primarily accounts for the $d$-orbitals, 
$\Ket{ d_{z^2}}$ and $\left(\Ket{d_{x^2-y^2}}\pm i\Ket{d_{xy}}\right)/\sqrt{2}$ \cite{xiao_coupled_2012}. The corresponding strained 
$k\cdot p$ Hamiltonian has been very recently suggested by Fang {\em et al.}, which in this basis attains the matrix form
\begin{equation}
H=\left[
\begin{array}{cc}
\left(f_0+\frac{f_1}{2}\right)+(f_3+f_4)(\varepsilon_{xx}+\varepsilon_{yy}) & f_2 a(k_x-ik_y)+f_5(\varepsilon_{xx}-\varepsilon_{yy}+2i\varepsilon_{xy}) \\
f_2 a(k_x+ik_y)+f_5(\varepsilon_{xx}-\varepsilon_{yy}-2i\varepsilon_{xy}) & \left(f_0-\frac{f_1}{2}\right)+(f_3-f_4)(\varepsilon_{xx}+\varepsilon_{yy}) 
\end{array}
\right]\, ,
\label{Hs}
\end{equation}
where $a$ is the lattice constant, $f_i$'s are the strained $k\cdot p$ parameters fitted to first-principles electronic bandstructure, 
also listed in Table~\ref{table-I} for convenience \cite{fang_electronic_2018}. Among these six $k\cdot p$ parameters, $f_0$ and $f_3$
do not play a role in homogeneous systems (as in this work). Their significance emerges in vertical heterostructures \cite{duan14} for $f_0$, 
and in localized strain gradients \cite{branny16, palacios-berraquero_large-scale_2017} for $f_3$. In Eq.~(\ref{Hs}) as well as in the 
remainder of this work, without loss of generality we refer to $K_+$ valley, which is assumed to be the origin for the wavevector 
$\vec{k}=\hat{x}k_x+\hat{y}k_y$, where $\hat{x}$ points from $\Gamma$ to $K_+$ in reciprocal space, that matches with the ``zigzag'' 
direction in direct space. Expressions for the $K_-$ valley, if required, can be obtained by complex conjugation, as the $K_\pm$ points 
are connected through time-reversal symmetry \cite{kormanyos13}.
The spin and the spin-orbit coupling are discarded in Eq.~(\ref{Hs}), although the spin-splitting can be easily 
incorporated to this framework \cite{fang_electronic_2018}. For our purposes this is not necessary as we are interested in the 
so-called $A$-excitons only \cite{conley_bandgap_2013}.

\begin{table}
\caption{$k.p$ parameters $f_i$ (eV), lattice constant $a$ (\AA) \cite{fang_electronic_2018}, 
2D polarizability $\chi_{\textit{2D}}$ (\AA)\cite{berkelbach_theory_2013} 
for different TMDs.}
\label{table-I} 
\small
\begin{center}
 \begin{tabular}{c| c| c| c| c| c| c| c| c} 
   Materials & $f_0$ & $f_1$  & $f_2$ & $f_3$ & $f_4$ &  $f_5$ & $a$ &  $\chi_{\textit{2D}}$ \\ [.5ex] 
     \hline
  MoS$_2$  & -5.07 & 1.79 & 1.06 & -5.47 & -2.59 & 2.2 & 3.182 & 6.60 \\ 
  MoSe$_2$ & -4.59 & 1.55 & 0.88 & -5.01 & -2.28 & 1.84 & 3.317 & 8.23 \\ 
  WS$_2$ & -4.66 & 1.95 & 1.22 & -5.82 & -3.59 & 2.27 & 3.182 & 6.03 \\ 
  WSe$_2$  & -4.23 & 1.65 & 1.02 & -5.26 & -3.02 & 2.03 & 3.316 & 7.18 \\ 
\end{tabular}
\end{center}
\end{table}

Neglecting any displacement perpendicular to TMD that lies on the two-dimensional (2D) $xy$-plane, the tensor strain components 
for most common types are:
\begin{eqnarray*}
\mathrm{Biaxial~strain:} & & \varepsilon_{yy}=\varepsilon_{xx}, ~~~ \varepsilon_{xy}=\varepsilon_{yx}=0,\\
\mathrm{Uniaxial~strain:} & & \varepsilon_{xx}\neq 0, ~~~ \varepsilon_{yy}=\varepsilon_{xy}=\varepsilon_{yx}=0,\\
\mathrm{Shear~strain:} & & \varepsilon_{yy}=-\varepsilon_{xx}, ~~~ \varepsilon_{xy}=\varepsilon_{yx}\neq 0,\\
\mathrm{Uniaxial~stress:} & & \varepsilon_{yy}=-\nu \varepsilon_{xx}, ~~~ \varepsilon_{xy}=\varepsilon_{yx}=0,
\end{eqnarray*}
where $\nu$ is the Poisson's ratio. To simplify our expressions, also we make use of 
the (areal) hydrostatic component of strain, $\varepsilon_H=\varepsilon_{xx}+\varepsilon_{yy}$.
Regarding the terminology, we should caution that the term uniaxial {\em strain} is in widespread use in TMD literature 
\cite{conley_bandgap_2013, roldan15, schmidt_reversible_2016, island_precise_2016}, although with the assumed Poisson's 
effect, as explicitly mentioned in these works, it needs to be referred to as uniaxial {\em stress}; also note that 
we use tensor and not the engineering strain \cite{nye85}.

The strained eigenstates of Eq.~(\ref{Hs}) can be readily solved analytically. The direct bandgap becomes 
$E_g = f_1+2f_4\varepsilon_H$. If we introduce 
$k_{x0} \equiv (\varepsilon_{xx}-\varepsilon_{yy})f_5/(f_2 a)$, $k_{y0} \equiv 2 \varepsilon_{xy}f_5/(f_2 a)$,
and $q_x\equiv k_x-k_{x0}$, $q_y\equiv k_y-k_{y0}$, with their magnitude $q=\sqrt{q_x^2+q_y^2}$, 
the energy dispersion for the conduction and valence bands can be expressed as
\begin{equation}
E_{c/v}(q)=f_0+f_3\varepsilon_H\pm \frac{E_g}{2} \sqrt{1+4\left[r(q,\varepsilon_H)\right]^2}, 
\label{cv}
\end{equation}
in terms of an auxiliary function that depends on the (valley edge-centered) wavenumber and the hydrostatic strain as
\begin{equation}
\label{rq}
r(q,\varepsilon_H) \equiv \frac{f_2aq}{f_1+2f_4 \varepsilon_H} ,
\end{equation}
which quantifies the degree of mixing between basis states $\left\{\Ket{\phi_c},\Ket{\phi_v}\right\}$ as to be justified below.

Hence, the valley edge shifts from $(k_x=0,k_y=0)$ to $(k_{x0},k_{y0})$ because of strain. 
So that for $\varepsilon_{xx}>\varepsilon_{yy}$, $k_{x0}>0$, and band extremum at $K$ shifts away from $\Gamma$ (toward neighboring 
zone $M^\prime$) point, while for $\varepsilon_{yy}>\varepsilon_{xx}$, $k_{x0}<0$ it shifts toward $\Gamma$ point. Thus, 
according to this simple model the shear strain rigidly displaces it along $k_y$ direction without affecting the bandgap. 
As a matter of fact the terms proportional to $f_5$ in Eq.~(\ref{Hs}) were referred to as the pseudogauge field in the graphene 
literature, responsible for shifting the Dirac cone from the $K$ point \cite{fang_electronic_2018}. 

The eigenvectors of the two-band Hamiltonian in Eq.~(\ref{Hs}) corresponding to the conduction and valence states are given by 
\begin{equation}
\label{states}
\ket{U_c}=\left( 
\begin{array}{c}
x_1 \\ x_2
\end{array}
\right),~~~ 
\ket{U_v}=\left( 
\begin{array}{c}
x_2 \\ -x_1^*
\end{array}
\right)\, , 
\end{equation}
where in terms of $\phi=\tan^{-1}(q_y/q_x)$, an $r$ defined in Eq.~(\ref{rq}) the entries are given by
\begin{eqnarray}
x_1 & = & \frac{e^{-i\phi}}{\sqrt{1+r^2}}, \\
x_2 & = & \frac{r}{\sqrt{1+r^2}}.
\end{eqnarray}

\subsection{Degree of circular polarization}
For a light polarized along a unit vector $\hat{u}$, the dipole matrix element connecting valence $\ket{U_v}$ and conduction $\ket{U_c}$
states is given by 
\cite{xiao_coupled_2012}
\begin{equation}
\label{CD}
\mathcal{P}_u (\vec{k}) \equiv \frac{m_0}{\hbar} \bra{U_c}\frac{\partial\hat{H}}{\partial k_u}\ket{U_v}\, ,
\end{equation}
where $m_0$ is the free electron mass.
For the $\pm$ circularly polarized light defined by the unit vectors
$\hat{u}_\pm=\left(\hat{x}\pm i \hat{y}\right)/\sqrt{2}$, 
Eq.~(\ref{CD}) can be expressed in terms of Pauli spin raising/lowering operators $\hat{\sigma}_\pm$ as
\begin{equation}
\mathcal{P}_\pm(\vec{k})=\frac{m_0 f_2 a}{\sqrt{2} \hbar} \braket{U_c|\hat{\sigma}_\pm|U_v}.
\end{equation}
Though it is not directly apparent from this expression, these momentum matrix elements are actually strain dependent 
through the eigenkets in Eq.~(\ref{states}) and the mixing function $r(q,\varepsilon_H)$ from Eq.~(\ref{rq}). 

The so-called circular dichroism (CD) corresponds to a difference in the absorption of the right- and left-hand circularly polarized radiation,
where the $k-$resolved degree of helicity selectivity is quantified as \cite{cao_valley-selective_2012},
\begin{equation}
\eta(\vec{k}) \equiv \frac{|\mathcal{P}_+(\vec{k})|^2-|\mathcal{P}_-(\vec{k})|^2}{|\mathcal{P}_+(\vec{k})|^2+|\mathcal{P}_-(\vec{k})|^2}\, .
\label{eta}
\end{equation}
We should note that inclusion of spin and the spin-orbit coupling does not affect the above helicity selection rules \cite{glazov15}.
For isotropic and electron-hole symmetric bands as in our case, its wavevector dependence simplifies as $\eta(\vec{k})\rightarrow \eta(q)=\eta(E)$. 
By inserting the associated states from Eq.~(\ref{states}) it acquires a simple analytical form 
\begin{equation}
\eta(q,\varepsilon_H)=\frac{1-\left[r(q,\varepsilon_H)\right]^4}{1+\left[r(q,\varepsilon_H)\right]^4},
\end{equation}
in terms of $r(q,\varepsilon_H)$ defined in Eq.~(\ref{rq}).

\subsection{Exciton binding and PL energies}
To compare with the experimental PL data under a given strain, we need to include excitonic effects as the associated 
binding energies significantly exceed the thermal energy at room temperature \cite{conley_bandgap_2013}. For that, we first 
extract the valley edge effective masses $m_{c/v}^{*}$ from the energy dispersion relation (Eq.~(\ref{cv})) via,
\begin{equation}
\label{m_cv}
m_{c/v}^{*}=\frac{\hbar^2}{\left(\left.\frac{\partial^2 E_{c/v}}{\partial k^2}\right|_{k_{x0},k_{y0}}\right)}
=\pm\frac{\hbar^2 \left(f_1+2f_4 \varepsilon_H\right)}{2(f_2 a)^2}\, ,
\end{equation}
where the curvatures are evaluated at the strained band extremum, $(k_{x0},k_{y0})$. 
Within our two-band $k\cdot p$ model electron and hole effective masses are equal, i.e., 
$m^*_e=m^*_c=m^*_h=-m^*_v$, and furthermore they are spatially isotropic,
but note that here these effective masses are strain-dependent. Thus, the corresponding exciton effective mass for its relative
degrees of freedom follows from $\mu=m^*_{e} m^*_{h}/\left(m^*_{e}+ m^*_{h}\right)$.

To retain the simplicity of our approach, the binding energies for neutral excitons in TMDs can be calculated following~\cite{berkelbach_theory_2013} by a variational method based on the exciton Hamiltonian (switching to Hartree atomic 
units in the remainder of this subsection),
\begin{equation}
H_{X}=-\frac{\nabla_{\textit{2D}}^{2}}{2\mu}+V_{\textit{2D}}(\rho)\, .
\end{equation}
Here, the in-plane interaction between an electron and a hole separated by $\rho=\sqrt{x^2+y^2}$ is,
\begin{equation}
V_{\textit{2D}}(\rho)=\frac{-\pi}{(\kappa_a+\kappa_b)\rho_0}\left[H_0(\rho/\rho_0)-Y_0(\rho/\rho_0)\right],
\end{equation}
where, $\kappa_a$ and $\kappa_b$ are dielectric constants for the media above and below the TMD (for a freestanding 
case, $\kappa_a=\kappa_b=1$), $H_0$ and $Y_0$ are the Struve and the Bessel function of the second kind; 
the screening length is given by $\rho_0 = 2\pi \chi_{\textit{2D}}$, and $\chi_{\textit{2D}}$ is the 2D polarizability of the 
TMD, which is listed in Table~\ref{table-I} \cite{berkelbach_theory_2013}.

The wave function for the neutral exciton with a single variational parameter $\lambda$ is chosen as,
\begin{equation}
\Psi_X(\rho;\lambda)=\sqrt{\frac{2}{\pi \lambda^2}}\exp{(-\rho/\lambda)}.
\end{equation}
In such a case the kinetic energy has the analytical form,
$
T(\lambda)=1/(2\mu \lambda^2),
$ 
while potential energy requires the following integration to be evaluated numerically, 
\begin{equation}
V(\lambda)=-\frac{2\pi}{\rho_0 \lambda^2} \int_{0}^{\infty} [H_0 (\rho/\rho_0)-Y_0 (\rho/\rho_0)]~\exp{(-2\rho/\lambda)}~\rho~d\rho.
\end{equation}
The total exciton energy is found by minimizing $E_X(\lambda) = T(\lambda)+V(\lambda)$, where the optimum value of $\lambda$ 
corresponds to the mean the exciton radius. For a bound exciton $E_X<0$, and the PL energy is obtained from $E_{\textit{PL}}=E_g+E_X$.

\subsection{A critique of the two-band Hamiltonian}
The Hamiltonian of Eq.~(\ref{Hs}) was originally derived from a simplified tight binding model for TMDs \cite{cazalilla14}.
Alternatively, starting from the well-known unstrained $k\cdot p$ model \cite{xiao_coupled_2012}, it can be arrived through the 
substitutions
\begin{equation}
\label{winkler-sub}
k_x\rightarrow k_x+\alpha\left(\varepsilon_{yy}-\varepsilon_{xx}\right)\, , ~~~ k_y\rightarrow k_y+\alpha 2\varepsilon_{xy}\, ,
\end{equation}
that resembles a minimal coupling to a strain-related gauge field with a coupling constant $\alpha$ (which is $f_5$ in our 
case) \cite{winkler10}. A group-theoretic basis of this substitution is that for the $C_{3h}$ point symmetry of the
$K$ point in TMDs, both $k_x-ik_y$ and $\left(\varepsilon_{xx}-\varepsilon_{yy}\right)+i2\varepsilon_{xy}$ transform according 
to $K_2$ ($\Gamma_3$ in the notation of \cite{koster63}), while $k_x+ik_y$ and $\left(\varepsilon_{xx}-\varepsilon_{yy}\right)-i2\varepsilon_{xy}$ 
transform according to $K_3$ ($\Gamma_2$) irreducible representations \cite{cheiwchanchamnangij13}. Moreover,
both $\Gamma_2$ and $\Gamma_3$ transform the same way under time-reversal symmetry \cite{koster63}.

Even though the substitution recipe in Eq.~(\ref{winkler-sub}) when applied to the unstrained Hamiltonian of \cite{xiao_coupled_2012}
generates strain terms respecting the symmetry of $K$ point, it fails to produce higher-order strain effects.
The rigorous approach following the method of invariants \cite{bir74} allows additional terms, like for instance,  
$\left(\varepsilon_{yy}-\varepsilon_{xx}\right)k_x +2\varepsilon_{xy}k_y$ on the diagonal entries of Eq.~(\ref{Hs}) \cite{winkler10}.
Unfortunately, at the present their coupling constants, similar to those in Table~1 are  unavailable.
In the absence of such terms, strain can only affect the dispersion via a single parameter $f_4$ (see, Eqs.~(\ref{cv}) and (\ref{rq}))
which amounts to a significant reduction in the degrees of freedom. As such, it is the underlying reason for the retention of 
isotropy of effective masses under the uniaxial deformation, 
which can be inferred from Eq.~(\ref{cv}) by the circular isoenergy curves for this case. However, from a quantitative point of view we 
believe that the implicit electron-hole symmetry from Eq.~(\ref{m_cv}) is more of a concern that equates their effective masses; and
yet, there exist other missing terms such as trigonal warping, and a cubic deviation in the band structure \cite{kormanyos13}.

\section{Results and discussions}
\subsection{Effect of strain on circular dichroism}
In pristine TMDs the CD stems from two crystal properties, namely the lack of center of inversion and the 
existance of the threefold rotational point symmetry, $C_3$ \cite{cao_valley-selective_2012}. At the $K_\pm$ point (i.e., $q=0$) CD is maximum ($\eta=1$)
that strictly allows the $\sigma^+$ ($\sigma^-$) helicity across the $K_+$ ($K_-$) direct transition between the two-band basis 
states of $\Ket{\phi_c}$ and $\Ket{\phi^\pm_v}$. Away from the $K$ point, the conduction and valence states develop an admixture of 
these basis states that leads to a degradation in the helicity discrimination, and hence in $\eta$. This can be mathematically followed from 
Eq.~(\ref{states}), where an increase in $q$ causes a mixing of the valley edge ($q=0$) states as mediated by the $r$ function (see, Eq.~(\ref{rq})). 

\begin{figure}[h!]
\centering\includegraphics[width=11cm]{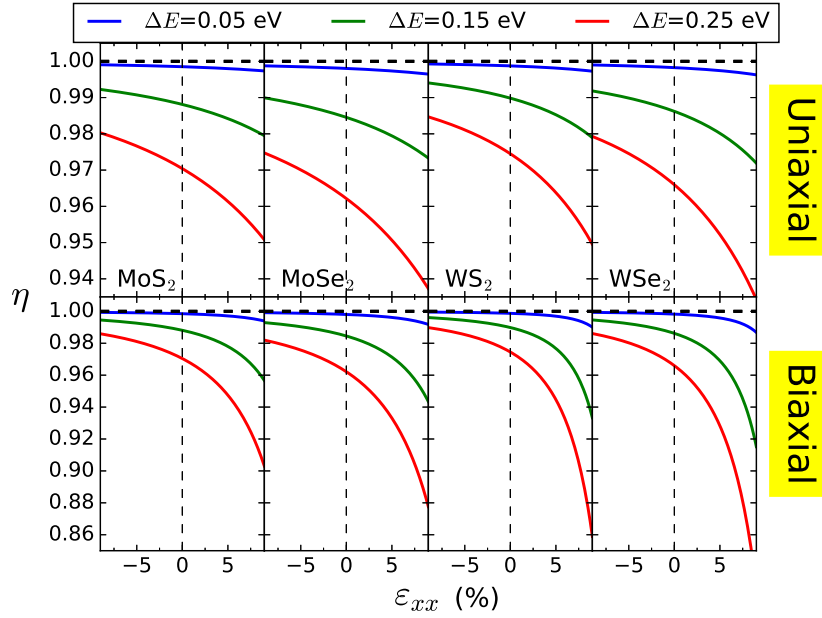}
\caption{\label{fig1} Effect of uniaxial/biaxial strain on the degree of optical polarization of TMDs for compressive/tensile strain at different 
excess energies $\Delta E$, as measured from the conduction band minimum.}
\end{figure}

To illustrate how strain affects this situation, using Eq.~(\ref{eta}) we plot in Fig.~\ref{fig1} $\eta(\Delta E)$ for various excess 
energies defined as $\Delta E=(E-E_g)/2$, where $E$ is the energy of the 
incoming photon.  The presence of a hydrostatic strain component $\varepsilon_H$ either enhances or
diminishes the variation in the mixing function $r$ and hence $\eta$ depending on the overall sign of $f_4\varepsilon_H$. 
With $f_4<0$ as seen from Table~1, this explains why the tensile strain $\varepsilon_H>0$ ($\varepsilon_{xx}>0$ in Fig.~1) inflates the 
variation in $\eta$ in Fig.~\ref{fig1}. We also observe that selenium-TMDs are more sensitive to strain in this respect, and the amount 
of change is larger for biaxial than uniaxial strain, as expected, for all of these materials. 

The above CD analysis is rather simplistic for a number of reasons. It does not include the excitonic interaction which would average 
$\eta(k)$ in a region of the $k$-space over which the exciton wave function extends. A more subtle consequence of the long range Coulomb interaction
between the electron and the hole is that it acts as an effective magnetic field causing the coupling of $\sigma^\pm$ polarizations of the exciton
that results in linearly polarized longitudinal and transverse eigenstates, and hence in the so-called linear dichroism \cite{glazov14,glazov15,wang18}. 
This is also left out of the scope of this work. Finally, any intervalley scattering \cite{mak12,kioseoglou12, aslan18} or other 
processes \cite{dery15} not considered here will further impair the CD. 

\subsection{PL peak shift under strain}
\begin{figure}[h!]
\centering\includegraphics[width=12cm]{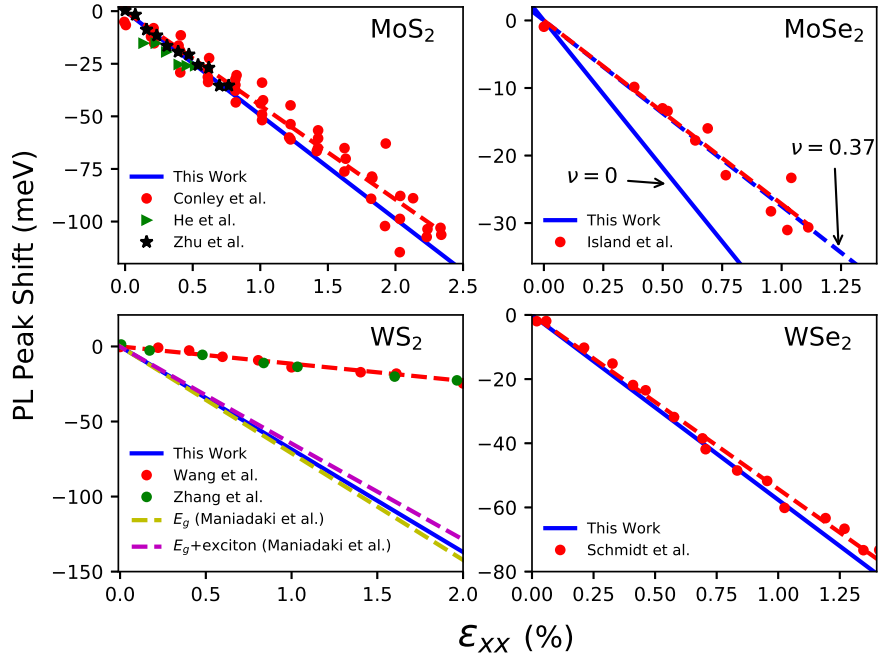}
\caption{\label{fig2} Uniaxial strain dependence of $A$-exciton PL peak energy shift for monolayer TMDs, comparing our calculations (in blue) 
with experimental data (symbols) along with their best fit line (red-dashed). References:  Conley {\em et al.}~\cite{conley_bandgap_2013}, 
Island {\em et al.}~\cite{island_precise_2016}, Wang {\em et al.}~\cite{wang_strain-induced_2015}, Zhang {\em et al.}~\cite{zhang16}, 
Schmidt {\em et al.}~\cite{schmidt_reversible_2016}, Maniadaki {\em et al.}~\cite{maniadaki_strain_2016}.}
\end{figure}

Figure~\ref{fig2} shows the PL peak shift for the four TMDs under uniaxial strain, comparing our calculations 
with the data from numerous experimental references. For MoS$_2$ and WSe$_2$ we have a good agreement between our theory and 
the best fit to the experimental data, 
taking into account the spread in the latter. At variance to this, for WS$_2$ our results do not 
agree with two reports \cite{wang_strain-induced_2015, zhang16}. To resolve this case, we also plot the bandgap 
variation for WS$_2$ under uniaxial strain from a first-principles calculation \cite{maniadaki_strain_2016} (yellow-dashed). 
If we add to this the strained excitonic correction we get a closer agreement with our calculations (purple-dotted vs. blue-solid). 
Therefore, we believe that some slipping may 
have occured on the TMD layer while applying strain to the substrate in~\cite{wang_strain-induced_2015, zhang16}, whereas 
other measurements in Fig.~\ref{fig2} such as~\cite{conley_bandgap_2013, schmidt_reversible_2016, island_precise_2016} have taken measures 
to clamp the TMD to the substrate. For MoSe$_2$, we see that the uniaxial {\em stress} condition using the Poisson's ratio of the substrate,
$\nu=0.37$ (blue-dashed) matches perfectly with the data, in agreement with their assertion in~\cite{island_precise_2016}. 
In other words, unlike the other measurements in Fig.~\ref{fig2}, for this experiment the TMD layer fully complies with the Poisson's contraction 
of the substrate.

\begin{table}[ht]
\caption{PL peak redshift under uniaxial or biaxial strain in comparison with results from literature.
Our results (this work) have both uniaxial strain/stress (i.e., $\nu :~$0/0.37) cases with the values in parentheses 
corresponding those without the excitonic contribution.}
\label{table-II} 
\small
\begin{center}
 \begin{tabular}{C{1.2cm}|C{1.3cm}|C{0.9cm}|C{1.3cm}|C{0.9cm}| C{1.3cm}| C{0.9cm}| C{1.3cm}| C{0.9cm}}
   \hline
   \multirow{2}{*}{meV${/\%}$} & \multicolumn{2}{c|}{MoS${_2}$} &   \multicolumn{2}{c|}{MoSe${_2}$} & \multicolumn{2}{c|}{WS${_2}$}& \multicolumn{2}{c}{WSe${_2}$} \\ \cline{2-9}
   & Uniaxial & Biaxial & Uniaxial & Biaxial & Uniaxial & Biaxial& Uniaxial & Biaxial \\        
   \hline 
 This Work & 49.4/31.1 (51.8/32.6) & 98.8 (103.6) & 43.6/27.5 (45.6/28.7) & 87.2 (91.2) & 68.5/43.2 (71.8/45.2) & 137 (143.6) & 57.6/36.3 (60.4/38.1) & 115.2 (120.8)\\
 \hline
  Literature & $\sim$45$^{a}$, $\sim$70$^{b}$, $\sim$48$^{c}$ & 99$\pm$6$^{d}$, (90.15), 90.1$^{e}$ & 27$\pm$2$^{f}$ & 53.74$^{e}$ & 11.3$^{g}$, 10$^{h}$ &  157$^{e}$& 54$^{i}$& 111$^{e}$\\ 
\hline  
\end{tabular}
\vspace{1ex}
\\
References: $^a$~\cite{conley_bandgap_2013}, $^b$~\cite{he_experimental_2013}, $^c$~\cite{zhu_strain_2013}, $^d$~\cite{lloyd_band_2016}, 
$^e$~\cite{frisenda_biaxial_2018}, $^f$~\cite{island_precise_2016}, $^g$~\cite{wang_strain-induced_2015}, 
$^h$~\cite{zhang16}, $^i$~\cite{schmidt_reversible_2016}
\end{center}
\end{table}
\begin{figure}[h!]
\centering\includegraphics[width=12cm]{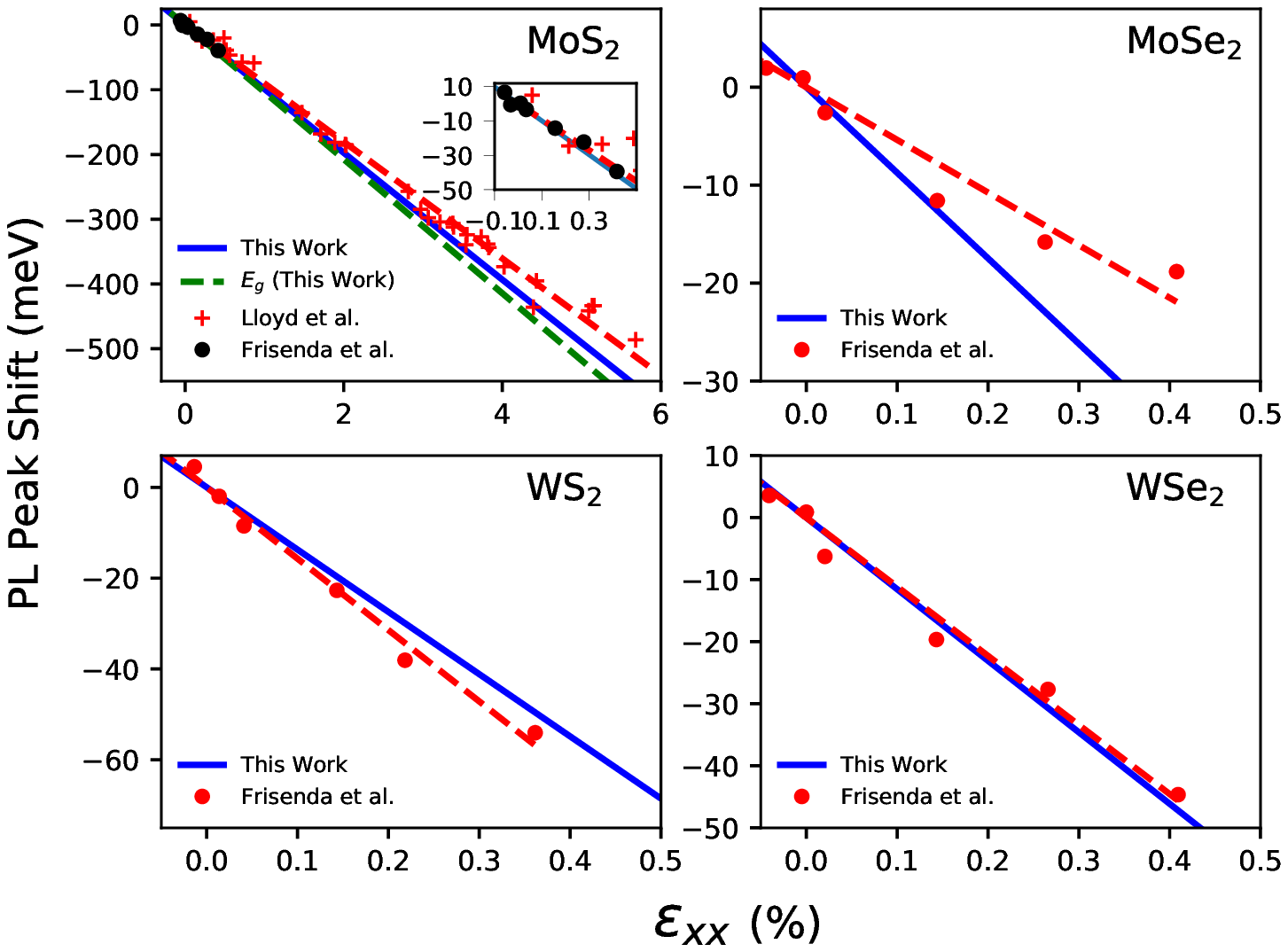}
\caption{\label{fig3} Biaxial strain dependence of $A$-exciton PL peak energy shift for monolayer TMDs, comparing our calculations (blue-solid) with experimental data 
(symbols) along with their best fit line (red-dashed). References: Lloyd {\em et al.}~\cite{lloyd_band_2016}, Frisenda {\em et al.}~\cite{frisenda_biaxial_2018}.}
\end{figure}

In the case of biaxial strain displayed in Fig.~\ref{fig3}, for MoS$_2$ we have an excellent agreement with the widest-range strain data 
by Lloyd {\em et al.} \cite{lloyd_band_2016} which goes up to 6\%. Once again we plot the variation in the bandgap 
(green-dashed in upper-left panel); 
the notable offset from PL line indicates the extend of excitonic contribution on the strain variation of the PL energy. 
Here, our results as well as~\cite{lloyd_band_2016} are for freestanding monolayer TMDs, on the other hand the 
remaining biaxial strain data from~\cite{frisenda_biaxial_2018} was originally reported with respect to polypropylene 
substrate strain. Therefore, to convert substrate 
strain results in~\cite{frisenda_biaxial_2018}, we multiplied all strain data from this reference by the $0.573$ scale factor. 
This brings their data in agreement with the freestanding case of~\cite{lloyd_band_2016}. However, for MoSe$_2$ we still 
have a disagreement with~\cite{frisenda_biaxial_2018}; noting the leveling off in their data beyond about 0.15\% strain, 
we again suspect that a slipping might be responsible. 

In Table~\ref{table-II} we compare our PL peak strain shift results with the quantities from various experimental references. 
For our results,  both uniaxial strain and uniaxial {\em stress} cases are presented, where in the former no transverse 
contraction takes place in the direction perpendicular to axial deformation (i.e., $\nu =0$). For uniaxial stress we use $\nu=0.37$ 
value which is typical for the flexible substrates in use \cite{schmidt_reversible_2016, island_precise_2016}.
As mentioned above, the uniaxial stress condition applies only for the MoSe$_2$ experiment of~\cite{island_precise_2016}. 
We also quote in parentheses our results excluding the variation of exciton binding energy under strain. It can be observed 
that sulfur-TMDs are more responsive to strain for PL peak shift and the amount of change is larger for biaxial strain 
than uniaxial one for each material considered.


\section{Conclusions}
A simple two-band $k\cdot p$ approach within minimal coupling to strain as applied to TMDs shows that
the bandgap and effective masses are affected by the hydrostatic component of the strain, whereas the shear 
strain does not alter the optoelectronic properties, but merely shifts the wavevector of the valley extrema.
A mixing of the valley edge states occurs away from the extrema which is either amplified or
diminished depending on the tensile or compressive nature of strain, respectively. This also manifests itself on the CD 
that can be tuned in either direction by applying tensile or compressive strain which is more pronounced for the biaxial case.
Comparison of strain-dependent PL peak shifts with a wide range of experimental data for monolayer TMDs demonstrates a satisfactory 
agreement provided that excitonic effects are included, and it reveals whether Poisson's effect takes place in a certain experiment. 
This analysis can easily be extended to other TMDs  with the availability of their $k\cdot p$ parameters. It can also act as a 
benchmark for more refined theories. 

\section*{Acknowledgments}
We are thankful to Dr. \"{O}zg\"{u}r Burak Aslan for fruitful discussions and suggestions.



\begin{thebibliography}{10}
\newcommand{\enquote}[1]{``#1''}

\bibitem{bertolazzi11}
S.~Bertolazzi, J.~Brivio, and A.~Kis, \enquote{Stretching and breaking of
  ultrathin {MoS}$_2$,} ACS Nano \textbf{5}, 9703--9709 (2011).

\bibitem{akinwande14}
D.~Akinwande, N.~Petrone, and J.~Hone, \enquote{Two-dimensional flexible
  nanoelectronics,} Nat. Commun. \textbf{5}, 5678 (2014).

\bibitem{roldan15}
R.~Rold\'{a}n, A.~Castellanos-Gomez, E.~Cappelluti, and F.~Guinea,
  \enquote{Strain engineering in semiconducting two-dimensional crystals,} J.
  Phys. :Condens. Matter \textbf{27}, 313201 (2015).

\bibitem{he_experimental_2013}
K.~He, C.~Poole, K.~F. Mak, and J.~Shan, \enquote{Experimental demonstration of
  continuous electronic structure tuning via strain in atomically thin
  {MoS}$_2$,} Nano Lett. \textbf{13}, 2931--2936 (2013).

\bibitem{conley_bandgap_2013}
H.~J. Conley, B.~Wang, J.~I. Ziegler, R.~F. Haglund, S.~T. Pantelides, and
  K.~I. Bolotin, \enquote{Bandgap engineering of strained monolayer and bilayer
  {MoS}$_2$,} Nano Lett. \textbf{13}, 3626--3630 (2013).

\bibitem{zhu_strain_2013}
C.~R. Zhu, G.~Wang, B.~L. Liu, X.~Marie, X.~F. Qiao, X.~Zhang, X.~X. Wu,
  H.~Fan, P.~H. Tan, T.~Amand, and B.~Urbaszek, \enquote{Strain tuning of
  optical emission energy and polarization in monolayer and bilayer {MoS}$_2$,}
  Phys. Rev. B \textbf{88}, 121301 (2013).

\bibitem{castellanos-gomez_local_2013}
A.~Castellanos-Gomez, R.~Rold\'{a}n, E.~Cappelluti, M.~Buscema, F.~Guinea,
  H.~S.~J. van~der Zant, and G.~A. Steele, \enquote{Local strain engineering in
  atomically thin {MoS}$_2$,} Nano Lett. \textbf{13}, 5361--5366 (2013).

\bibitem{tonndorf13}
P.~Tonndorf, R.~Schmidt, P.~B\"{o}ttger, X.~Zhang, J.~B\"{o}rner, A.~Liebig,
  M.~Albrecht, C.~Kloc, O.~Gordan, D.~R.~T. Zahn, S.~M. de~Vasconcellos, and
  R.~Bratschitsch, \enquote{Photoluminescence emission and raman response of
  monolayer {MoS}$_2$, {MoSe}$_2$, and {WSe}$_2$,} Opt. Express \textbf{21},
  4908--4916 (2013).

\bibitem{hui_exceptional_2013}
Y.~Y. Hui, X.~Liu, W.~Jie, N.~Y. Chan, J.~Hao, Y.-T. Hsu, L.-J. Li, W.~Guo, and
  S.~P. Lau, \enquote{Exceptional tunability of band energy in a compressively
  strained trilayer {MoS}$_2$ sheet,} ACS Nano \textbf{7}, 7126--7131 (2013).

\bibitem{sercombe13}
D.~Sercombe, S.~Schwarz, O.~Del Pozo-Zamudio, F.~Liu, B.~J. Robinson, E.~A.
  Chekhovich, I.~I. Tartakovskii, O.~Kolosov, and A.~I. Tartakovskii,
  \enquote{Optical investigation of the natural electron doping in thin
  {MoS}$_2$ films deposited on dielectric substrates,} Sci. Rep. \textbf{3},
  3489 (2013).

\bibitem{lloyd_band_2016}
D.~Lloyd, X.~Liu, J.~W. Christopher, L.~Cantley, A.~Wadehra, B.~L. Kim, B.~B.
  Goldberg, A.~K. Swan, and J.~S. Bunch, \enquote{Band gap engineering with
  ultralarge biaxial strains in suspended monolayer {MoS}$_2$,} Nano Lett.
  \textbf{16}, 5836--5841 (2016).

\bibitem{branny16}
A.~Branny, G.~Wang, S.~Kumar, C.~Robert, B.~Lassagne, X.~Marie, B.~D. Gerardot,
  and B.~Urbaszek, \enquote{Discrete quantum dot like emitters in monolayer
  {MoSe}$_2$: Spatial mapping, magneto-optics, and charge tuning,} Appl. Phys.
  Lett. \textbf{108}, 142101 (2016).

\bibitem{palacios-berraquero_large-scale_2017}
C.~Palacios-Berraquero, D.~M. Kara, A.~R.-P. Montblanch, M.~Barbone,
  P.~Latawiec, D.~Yoon, A.~K. Ott, M.~Loncar, A.~C. Ferrari, and
  M.~Atat\"{u}re, \enquote{Large-scale quantum-emitter arrays in atomically
  thin semiconductors,} Nat. Commun. \textbf{8}, 15093 (2017).

\bibitem{plechinger_control_2015}
G.~Plechinger, A.~Castellanos-Gomez, M.~Buscema, H.~S. J. v.~d. Zant, G.~A.
  Steele, A.~Kuc, T.~Heine, C.~Sch\"{u}ller, and T.~Korn, \enquote{Control of
  biaxial strain in single-layer molybdenite using local thermal expansion of
  the substrate,} 2D Mater. \textbf{2}, 015006 (2015).

\bibitem{li_optoelectronic_2015}
H.~Li, A.~W. Contryman, X.~Qian, S.~M. Ardakani, Y.~Gong, X.~Wang, J.~M.
  Weisse, C.~H. Lee, J.~Zhao, P.~M. Ajayan, J.~Li, H.~C. Manoharan, and
  X.~Zheng, \enquote{Optoelectronic crystal of artificial atoms in
  strain-textured molybdenum disulphide,} Nat. Commun. \textbf{6}, 7381 (2015).

\bibitem{desai_strain-induced_2014}
S.~B. Desai, G.~Seol, J.~S. Kang, H.~Fang, C.~Battaglia, R.~Kapadia, J.~W.
  Ager, J.~Guo, and A.~Javey, \enquote{Strain-induced indirect to direct
  bandgap transition in multilayer {WSe}$_2$,} Nano Lett. \textbf{14},
  4592--4597 (2014).

\bibitem{yang_tuning_2015}
S.~Yang, C.~Wang, H.~Sahin, H.~Chen, Y.~Li, S.-S. Li, A.~Suslu, F.~M. Peeters,
  Q.~Liu, J.~Li, and S.~Tongay, \enquote{Tuning the optical, magnetic, and
  electrical properties of {ReSe}$_2$ by nanoscale strain engineering,} Nano
  Lett. \textbf{15}, 1660--1666 (2015).

\bibitem{wang_strain-induced_2015}
Y.~Wang, C.~Cong, W.~Yang, J.~Shang, N.~Peimyoo, Y.~Chen, J.~Kang, J.~Wang,
  W.~Huang, and T.~Yu, \enquote{Strain-induced direct-indirect bandgap
  transition and phonon modulation in monolayer {WS}$_2$,} Nano Res.
  \textbf{8}, 2562--2572 (2015).

\bibitem{zhang16}
Q.~Zhang, Z.~Chang, G.~Xu, Z.~Wang, Y.~Zhang, Z.-Q. Xu, S.~Chen, Q.~Bao, J.~Z.
  Liu, Y.-W. Mai \emph{et~al.}, \enquote{Strain relaxation of monolayer
  {WS}$_2$ on plastic substrate,} Adv. Funct. Mater. \textbf{26}, 8707--8714
  (2016).

\bibitem{schmidt_reversible_2016}
R.~Schmidt, I.~Niehues, R.~Schneider, M.~Dr\"{u}ppel, T.~Deilmann, {Michael
  Rohlfing}, S.~M.~d. Vasconcellos, A.~Castellanos-Gomez, and R.~Bratschitsch,
  \enquote{Reversible uniaxial strain tuning in atomically thin {WSe}$_2$,} 2D
  Mater. \textbf{3}, 021011 (2016).

\bibitem{island_precise_2016}
J.~O. Island, A.~Kuc, E.~H. Diependaal, R.~Bratschitsch, H.~S. J. v.~d. Zant,
  T.~Heine, and A.~Castellanos-Gomez, \enquote{Precise and reversible band gap
  tuning in single-layer {MoSe}$_2$ by uniaxial strain,} Nanoscale \textbf{8},
  2589--2593 (2016).

\bibitem{peelaers12}
H.~Peelaers and C.~G. Van~de Walle, \enquote{Effects of strain on band
  structure and effective masses in {MoS}$_2$,} Phys. Rev. B \textbf{86},
  241401 (2012).

\bibitem{rostami_theory_2015}
H.~Rostami, R.~Rold\'{a}n, E.~Cappelluti, R.~Asgari, and F.~Guinea,
  \enquote{Theory of strain in single-layer transition metal dichalcogenides,}
  Phy. Rev. B \textbf{92}, 195402 (2015).

\bibitem{maniadaki_strain_2016}
A.~E. Maniadaki, G.~Kopidakis, and I.~N. Remediakis, \enquote{Strain
  engineering of electronic properties of transition metal dichalcogenide
  monolayers,} Solid State Commun. \textbf{227}, 33--39 (2016).

\bibitem{feierabend_impact_2017}
M.~Feierabend, A.~Morlet, G.~Bergh\"{a}user, and E.~Malic, \enquote{Impact of
  strain on the optical fingerprint of monolayer transition-metal
  dichalcogenides,} Phys. Rev. B \textbf{96}, 045425 (2017).

\bibitem{cao_valley-selective_2012}
T.~Cao, G.~Wang, W.~Han, H.~Ye, C.~Zhu, J.~Shi, Q.~Niu, P.~Tan, E.~Wang,
  B.~Liu, and J.~Feng, \enquote{Valley-selective circular dichroism of
  monolayer molybdenum disulphide,} Nat. Commun. \textbf{3}, 887 (2012).

\bibitem{xiao_coupled_2012}
D.~Xiao, G.-B. Liu, W.~Feng, X.~Xu, and W.~Yao, \enquote{Coupled spin and
  valley physics in monolayers of {MoS}$_2$ and other group-{VI}
  dichalcogenides,} Phys. Rev. Lett. \textbf{108}, 196802 (2012).

\bibitem{fang_electronic_2018}
S.~Fang, S.~Carr, M.~A. Cazalilla, and E.~Kaxiras, \enquote{Electronic
  structure theory of strained two-dimensional materials with hexagonal
  symmetry,} Phys. Rev. B \textbf{98}, 075106 (2018).

\bibitem{duan14}
X.~Duan, C.~Wang, J.~C. Shaw, R.~Cheng, Y.~Chen, H.~Li, X.~Wu, Y.~Tang,
  Q.~Zhang, A.~Pan \emph{et~al.}, \enquote{Lateral epitaxial growth of
  two-dimensional layered semiconductor heterojunctions,} Nat. Nanotech.
  \textbf{9}, 1024 (2014).

\bibitem{kormanyos13}
A.~Korm\'anyos, V.~Z\'olyomi, N.~D. Drummond, P.~Rakyta, G.~Burkard, and V.~I.
  Falko, \enquote{Monolayer {MoS}$_{2}$: Trigonal warping, the
  $\ensuremath{\Gamma}$ valley, and spin-orbit coupling effects,} Phys. Rev. B
  \textbf{88}, 045416 (2013).

\bibitem{berkelbach_theory_2013}
T.~C. Berkelbach, M.~S. Hybertsen, and D.~R. Reichman, \enquote{Theory of
  neutral and charged excitons in monolayer transition metal dichalcogenides,}
  Phys. Rev. B \textbf{88}, 045318 (2013).

\bibitem{nye85}
J.~F. Nye, \emph{Physical properties of crystals: their representation by
  tensors and matrices} (Oxford University Press, 1985).

\bibitem{glazov15}
M.~M. Glazov, E.~L. Ivchenko, G.~Wang, T.~Amand, X.~Marie, B.~Urbaszek, and
  B.~L. Liu, \enquote{Spin and valley dynamics of excitons in transition metal
  dichalcogenide monolayers,} Phys. Status Solidi B \textbf{252}, 2349--2362
  (2015).

\bibitem{cazalilla14}
M.~Cazalilla, H.~Ochoa, and F.~Guinea, \enquote{Quantum spin hall effect in
  two-dimensional crystals of transition-metal dichalcogenides,} Phys. Rev.
  Lett. \textbf{113}, 077201 (2014).

\bibitem{winkler10}
R.~Winkler and U.~Z\"ulicke, \enquote{Invariant expansion for the trigonal band
  structure of graphene,} Phys. Rev. B \textbf{82}, 245313 (2010).

\bibitem{koster63}
G.~F. Koster, J.~D. Dimmock, R.~G. Wheeler, and H.~Statz, \emph{Properties of
  the thirty-two point groups} (MIT Press, 1963).

\bibitem{cheiwchanchamnangij13}
T.~Cheiwchanchamnangij, W.~R.~L. Lambrecht, Y.~Song, and H.~Dery,
  \enquote{Strain effects on the spin-orbit-induced band structure splittings
  in monolayer {MoS}$_2$ and graphene,} Phys. Rev. B \textbf{88}, 155404
  (2013).

\bibitem{bir74}
G.~L. Bir and G.~E. Pikus, \emph{Symmetry and strain-induced effects in
  semiconductors} (Wiley/Halsted Press, 1974).

\bibitem{glazov14}
M.~M. Glazov, T.~Amand, X.~Marie, D.~Lagarde, L.~Bouet, and B.~Urbaszek,
  \enquote{Exciton fine structure and spin decoherence in monolayers of
  transition metal dichalcogenides,} Phys. Rev. B \textbf{89}, 201302 (2014).

\bibitem{wang18}
G.~Wang, A.~Chernikov, M.~M. Glazov, T.~F. Heinz, X.~Marie, T.~Amand, and
  B.~Urbaszek, \enquote{Colloquium: {Excitons} in atomically thin transition
  metal dichalcogenides,} Rev. Mod. Phys. \textbf{90}, 021001 (2018).

\bibitem{mak12}
K.~F. Mak, K.~He, J.~Shan, and T.~F. Heinz, \enquote{Control of valley
  polarization in monolayer {MoS}$_2$ by optical helicity,} Nat. Nanotech.
  \textbf{7}, 494 (2012).

\bibitem{kioseoglou12}
G.~Kioseoglou, A.~T. Hanbicki, M.~Currie, A.~L. Friedman, D.~Gunlycke, and
  B.~T. Jonker, \enquote{Valley polarization and intervalley scattering in
  monolayer {MoS}$_2$,} Appl. Phys. Lett. \textbf{101}, 221907 (2012).

\bibitem{aslan18}
O.~B. Aslan, I.~M. Datye, M.~J. Mleczko, K.~Sze~Cheung, S.~Krylyuk, A.~Bruma,
  I.~Kalish, A.~V. Davydov, E.~Pop, and T.~F. Heinz, \enquote{Probing the
  optical properties and strain-tuning of ultrathin
  {Mo}$_{1-x}${W}$_x${Te}$_2$,} Nano Lett. \textbf{18}, 2485--2491 (2018).

\bibitem{dery15}
H.~Dery and Y.~Song, \enquote{Polarization analysis of excitons in monolayer
  and bilayer transition-metal dichalcogenides,} Phys. Rev. B \textbf{92},
  125431 (2015).

\bibitem{frisenda_biaxial_2018}
R.~Frisenda, R.~Schmidt, S.~M. de~Vasconcellos, R.~Bratschitsch, D.~P. de~Lara,
  and A.~Castellanos-Gomez, \enquote{Biaxial strain in atomically thin
  transition metal dichalcogenides,} arXiv:1804.11095 [cond-mat]  (2018).
  ArXiv: 1804.11095.

\end{thebibliography}
\end{document}